\documentclass[sigconf, nonacm]{acmart}
\usepackage[utf8]{inputenc}
\usepackage{graphicx}
\graphicspath{ {Figures/} }
\usepackage{placeins}
\usepackage{booktabs}
\usepackage{wrapfig}

\newcommand{\vknote}[1]{\textcolor{teal}{VK: #1}}
\renewcommand{\vknote}[1]{}

\newcommand{\omt}[1]{}

\title{Exporting Geography Into A Virtual Landscape: A Global Pandemic Locally Discussed}
\author{Katherine Van Koevering}
\email{kav64@cornell.edu}
\affiliation{
  \institution{Cornell University}
  \city{Ithaca}
  \state{New York}
  \country{USA}
}

\author{Yiquan Hong}
\email{yh846@cornell.edu}
\affiliation{
  \institution{Cornell University}
  \city{Ithaca}
  \state{New York}
  \country{USA}
}

\author{Jon Kleinberg}
\email{kleinberg@cornell.edu}
\affiliation{
  \institution{Cornell University}
  \city{Ithaca}
  \state{New York}
  \country{USA}
}
\setcopyright{none}
\begin{document}

\begin{abstract}
    % The COVID-19 pandemic is perhaps unique in that it is the first global health crisis to happen in the age of social media. Due to the nature of the crisis, much of every day social interaction was forced online, providing an enormous record of reactions and conversations surrounding the impact. However, while social media allows for interaction regardless of location, the pandemic itself is inherently tied to geography, giving a glimpse into how real-world geography translates onto a virtual landscape. By analyzing almost 300 geographically-linked COVID discussion communities on Reddit, we show how these discussions were influenced geographically and temporally in three aspects: what were people talking about, who were they talking about it with, and how did they self-organize these conversations?

    The COVID-19 pandemic has been a global health crisis playing out in the age of social media.  Even though the virtual environment makes interaction possible regardless of physical location, many of the most pressing issues during the pandemic --- case counts, lockdown policies, vaccine availability --- have played out in an intensely local fashion. Reflecting this locality, many of the online COVID communities that formed have been closely tied to physical location, at different spatial scales ranging from cities to countries to entire global platforms. This provides an opportunity to study how the real-world geography of the pandemic translates into a virtual landscape. By analyzing almost 300 geographically-linked COVID discussion communities on Reddit, we show how these discussions were organized geographically and temporally in three aspects: what were people talking about, who were they talking about it with, and how did they self-organize these conversations?

    % Due to the nature of the crisis, much of every day social interaction was forced online, providing an enormous record of reactions and conversations surrounding the impact. However, while social media allows for interaction regardless of location, the pandemic itself is inherently tied to geography, giving a glimpse into how real-world geography translates onto a virtual landscape. By analyzing almost 300 geographically-linked COVID discussion communities on Reddit, we show how these discussions were influenced geographically and temporally in three aspects: what were people talking about, who were they talking about it with, and how did they self-organize these conversations?
    
\end{abstract}

\maketitle

\section{Introduction}

The COVID-19 pandemic has been a global subject of intense focus on social media since its early stages \cite{gozzi2020collective,zhang-covid-subreddits}, and it has exemplified many of the fundamental properties of social media usage during major world events. People have used social media for social support \cite{andy2021understanding} and online sense-making, both of which are common practices during unfolding crises \cite{nrc-alerts-social-media,vieweg2010microblogging}. Online discussion about COVID has had strong political dimensions \cite{rajadesingan2021political}, and social media has been a source of COVID misinformation and conspiracy theories \cite{ashford2022understanding,dow2021covid}. 

For most people, COVID has been both a global phenomenon and a local one: many of the most pressing issues, including case counts, lockdown policies, vaccine availability, the closing and reopening of businesses and schools, and many other considerations were fundamentally local in nature; they varied considerably from one place to another. This locality manifested itself in social media as well, with online communities forming to focus on COVID in the context of specific cities, states, and countries. Such specialization is consistent with the ways in which online topics acquire strong geographic locality \cite{gravano2003categorizing,wang2005detecting,backstrom2008spatial} more generally: despite the virtual nature of the medium, which makes communication possible in principle across large distances, the topics themselves are often of the most interest to people in a constrained geographic region. 

In the case of COVID, this geographic specialization makes it possible to investigate questions about geographic variation in response to the pandemic. To begin with, we know through news reporting that different parts of the world, and different regions within large countries including the United States, had very different reactions to the pandemic, and appeared to develop different norms around the ways in which authoritative public health recommendations were handled. Can we see this reflected in differences among the discussions and the types of information shared in online COVID communities associated with different geographic locations? 

Geographic specialization of communities also enables us to ask questions not just about the pandemic, but as a case study in how online media self-organizes at different geographic scales in response to rapidly-evolving global events. How was this geographic differentiation initiated online, and how early into the pandemic did it occur? How did it balance between ``horizontal'' specialization across similarly-sized regions and ``vertical'' specialization into a hierarchy of communities for cities, regions, countries, and the entire world? And did users tend to participate broadly across multiple geographically-focused communities over time, or did they remain focused on one or very few? 

\paragraph*{\bf The present work: COVID subreddits organized geographically} Here we address these questions using a large-scale dataset of COVID communities drawn from Reddit (reddit.com). Reddit has been one of the major loci of online COVID discussion, and its communities ({\em subreddits}, in the terminology of Reddit) exhibit the properties discussed above at a broad scale. Early in the pandemic, COVID subreddits formed both for the pandemic in general, as well as for a large number of countries, for each of the 50 US states, and for many large cities. We have assembled a dataset containing all of these subreddits, collectively amounting to a significant proportion of the COVID discussion taking place on Reddit.
We consider questions at both global and local scales, with particular analysis of US geography because of the availability of subreddits for all 50 US states.

%To address the questions discussed above, we organize our analysis of these COVID subreddits into three general themes. We first consider the information shared on the subreddits: how did these vary across geography, and did the sharing of information associated with authoritative public health sources --- for examples links to CDC Web sites --- vary according to patterns of political difference between the geographic regions? We then consider the set of users and how they divided themselves among the subreddits, considering both the typical number of subreddits that users contributed to, and also the geographic locality in these patterns. Finally, we consider the self-organization of the subreddits themselves, including the dynamics by which they originated, and some of interesting consequences of the rapid formation of the communities, including the presence of {\em isomorphic} communities --- different subreddits devoted to the same local geographic region. 

To address the questions discussed above, we organize our analysis of these COVID subreddits into three general themes: content, users, and communities. We first consider the information shared on the subreddits: how did these vary across geography, and did the sharing of information associated with authoritative public health sources --- for examples links to CDC Web sites --- vary according to patterns of political difference between the geographic regions? This analysis reveals several themes among URL usage. One theme is that, using vote shares for the two main US political parties in the 2020 US Presidential election,  US states with Democratic majorities tend to share more common domains, such as medxriv.org, than Republican states, which share more uncommon domains, such as thetennessean.com. Additionally, many of the domains that are most overrepresented in states with democractic majorities correspond to scientific authorities, an interesting and fundamental social-media reflection of a partisan phenomenon that has appeared in survey data \cite{lee2021party}.

We then consider the set of users and how they divided themselves among the subreddits, considering both the typical number of subreddits that users contributed to, and also the geographic locality in these patterns. One common theme among users is that users who participate in more subreddits, rather than spreading their activity more thinly among this large collection of subreddits, are much more active on a per-subreddit basis. In particular, for a user that participates in $n$ subreddits, the amount of activity in their $k^{\rm th}$ favorite subreddit will on average be higher than the amount of activity in the $(k-1)^{\rm st}$ favorite subreddit for a user with $n-1$ subreddits. However, the average amount of activity in a user's favorite community stands at approximately 50\% of their total activity regardless of how many communities a user is part of. Another theme is that users have a tendency to broaden their community diversity vertically before horizontally, seeking communities at different geographic scales rather than at the same scale.

Finally, we consider the self-organization of the subreddits themselves, including the dynamics by which they originated, and some of interesting consequences of the rapid formation of the communities, including the presence of {\em isomorphic} communities --- different subreddits devoted to the same local geographic region. One theme is that subreddit creation is remarkably similar for nearly all subreddits. Nearly all communities we include were created within a four week period around early March, 2020, although subreddits for larger geographic regions tended to be created slightly earlier than those for smaller geographic regions. Additionally, the users who created these communities tended to stay very involved, even in relatively large communities, often contributing a significant amount of content. However, in cases where multiple subreddits are created for the same location, there is nearly always one dominant subreddit, and that is usually the first subreddit created.

\subsection{Data}
For our analysis, we use 281 COVID subreddits. These were found by identifying all subreddits with the words `coronavirus' or `covid' in the name and manually filtering for subreddits that designated a location and were general COVID subreddits (for instance, a lockdown specific subreddit would be excluded). Subreddits that had minimal activity were also excluded. Based on additional followup investigations, this is nearly all relevant subreddits.

These subreddits were then manually tagged with the location the represented, and given a \textbf{level} of either 'top', 'country', 'state', 'city', 'state\_f' or 'city\_f' to designate the vertical level of the location. These represented general (non-geographic or global) subreddits, country-specific subreddits, US state subreddits, US city subreddits, non-US state subreddits, and non-US city subreddits, respectively. In the case that a subreddit was ambiguous (e.x. Southern California), the level closest to the size of the region was specified. In this case, Southern California is considered state level. A European subreddit would be considered country level, since it is closer to a country than a top-level general or global subreddit. Additionally, for most analysis, suspended and banned users (and their activities) were removed from the data set.

\subsection{Overview of Related Work}

Our paper is related to several branches of literature.
Most specifically, it follows up on recent efforts to study the impact of COVID on social media by focusing on the geographic dimensions of these online communities and discussions.
In work that takes a more global view, focusing on COVID communities independent of geography, Zhang et al \cite{zhang-covid-subreddits} study the organization and evolution of the global COVID subreddits, which we also include as the top level of our geographic hierarchy.
Ashford et al and Dow et al \cite{ashford2022understanding,dow2021covid} consider misinformation and conspiracies about COVID online, though again the focus is not principally geographic.
In work that is more similar to ours in its geographic focus, Yan et al use Reddit to study the sentiment about vaccines in particular across cities in Canada \cite{yan2021comparing}; we broaden the questions to a larger enumeration of topics and locations, and at multiple spatial scales. Aubin Le Qu\'{e}r\'{e} et al focus on the local news coverage of COVID on online communities, but do not analyze the communities \cite{lelocal} themselves.

Generalizing beyond COVID, our work is also related to the study of social media use during crises \cite{nrc-alerts-social-media,vieweg2010microblogging}.
In particular, the type of self-organization and collective sense-making through online interaction has been described in earlier work by Starbird et al as a kind of ``verbal milling behavior'' \cite{starbird2016could} to seek out and share information in the midst of uncertainty.

Our work is also related to general efforts to map the spatial organization of different forms of online information. Early work looked at the geographic loci of search engine queries \cite{gravano2003categorizing,wang2005detecting,backstrom2008spatial}, and recent papers have studied the effectiveness of online activity at identifying hot-spots of COVID \cite{fulk2022using,hisada2020surveillance,mcdonald2021can,yabe2022early}. 

\section{Content}
We focus our analysis of community content on two particular items: URLs shared by users, and keywords. These two items provide key insight into the topics being discussed and the sources of information relied upon for those discussions. These items reflect the theme of political differences, particularly with respect to what URLs are shared in different states. It also reflects differences in topics and outside information sources both vertically and horizontally across geographic locations.

\subsection{URLs}
URLs are a unique opportunity to look at where users were drawing outside information from and how they are relating their community discussions to the larger internet. Particularly in a global emergency such as COVID, where information was sparse and spread rapidly, these URLs are an important part of understanding what shaped the flow of discussion and who users referenced and trusted.

We start by asking which URLs are commonly shared. Unsurprisingly, the vast majority of these are COVID related (Fig.~\ref{fig:url_tree}). The top two categories are vaccine related and COVID data related, accounting for over half of the 200 most popular URLs. This is likely partially because there are regional variations among these two categories, resulting in many URLs. The fact that these URLs are still among the most popular despite this division, suggests that the communities were frequently used for sharing current information on both COVID rates and vaccine information, making them sort of community bulletin boards. Several subreddits scheduled weekly or daily posts of updated local COVID rates for part of the pandemic. 

\begin{figure}
    \centering
    \includegraphics[width=0.2\textwidth]{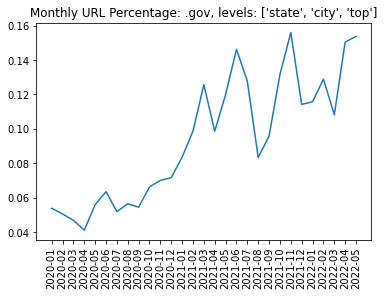}
    \includegraphics[width=0.2\textwidth]{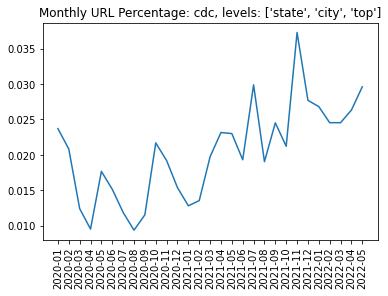}
    \includegraphics[width=0.2\textwidth]{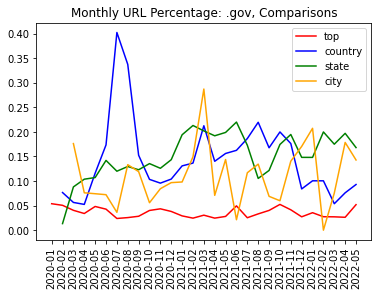}
    \includegraphics[width=0.2\textwidth]{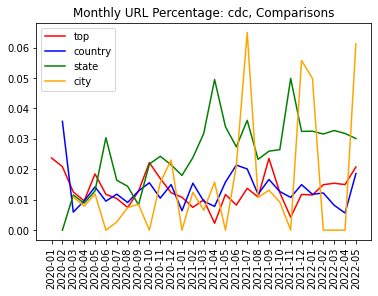}
    
    \caption{Comparisons of the prevalence of URLs containing '.gov' and 'cdc'. Top row for the city, state and top levels only (to filter for mostly US-based content). The bottom row for all levels compared.}
    \label{fig:gov_cdc_comp}
\end{figure}

Of course, what URLs are popular is not static (Fig.~\ref{fig:gov_cdc_comp}).  For instance, let us look at the temporal aspects of .gov URLs and CDC URLs. Both are common occurrences throughout the pandemic, but we can see a rise in their usage as the pandemic continues - which is perhaps surprising given that one might have expected authoritative sources such as government sites to be most popular at the beginning of the pandemic. This rise is not smooth, however, and if we break down the usage into levels we can see that there is quite a lot of noise. While top levels seem to have a relatively consistent level of about 1-3\% .gov and CDC URLs (much lower than other levels, perhaps due to a larger number of interactions overall), lower levels seem more reactive. This is partially due to less data in these levels.

\omt{
\begin{figure}
    \centering
    \includegraphics[width=0.2\textwidth]{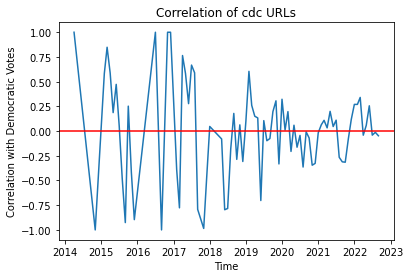}
    \includegraphics[width=0.2\textwidth]{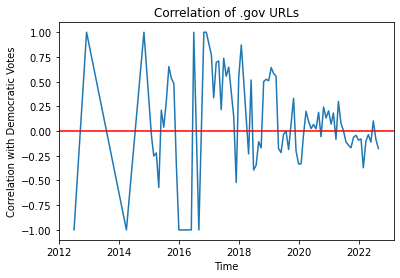}
    \caption{The political correlation of URLs containing 'cdc' on a monthly basis across all general state subreddits.}
    \label{fig:my_label}
\end{figure}
}

\begin{figure}[tb]
    \centering
    \includegraphics[width=0.45\textwidth]{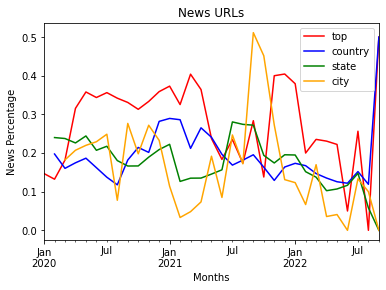}
    \caption{The percentage of all URLs that lead to a news website over time across levels.}
    \label{fig:news_domains}
\end{figure}

Another major component of URLs in the pandemic, beyond just government sites, was outside news sources. As such, we filtered for news domains \cite{bernhardclemm_2021} and analyzed the popularity of these domains (Fig.~\ref{fig:news_domains}). Overall, news domains remained popular throughout the pandemic, fulfilling the subreddits' role as community bulletin boards. However, popularity did shift over time. For instance, among top level subreddits,  we see a spike in news URLs at the beginning of the pandemic and around the delta spike for cities and states, but for top levels subreddits we see a spike near the omicron wave. It's possible that this reflects a more localized desire for updates for the Delta wave than Omicron wave.

News sources are, particularly during the pandemic, inherently political. So it is worth asking if that political bent is evident in the URLs used in these online discussions. We test this by checking the correlation between the popularity of a URL in a given state and the percentage of the state that voted for Biden in the 2020 presidential election. This can then be compared to a null model in which we randomize the states in which URLs appeared. Checking the 200 most popular domains among state subreddits, we find 15 domains with strong correlation with Democratic votes and only one domain (thetennessean.com) with strong correlation with Republican votes (Table~\ref{tab:dem_corr}). This suggests that Democratic states tend to use more popular domains more frequently than Republican states. We confirm this by checking to see how many domains it takes to account for random samples of the URLs used in different states. Democratic states generally have fewer domains than Republican states (average $\sim 350$ versus an average of $\sim 400$), suggesting more variation of the domains in use in Republican-leaning states. This could reflect a "mainstream" versus "alternative" influence.

% \begin{figure}
%     \centering
%     \includegraphics[width=0.4\textwidth]{url_samples.png}
%     \caption{100 trials of sampling 2000 instances of URL usage each for Republican-leaning and Democratic-leaning states. We then histogram how many domains it takes to account for half of these URLs.}
%     \label{fig:url_samples}
% \end{figure}

\begin{wraptable}{r}{4.5cm}
\centering
\caption{Correlation of domain prevalence with Democratic votes for domains with p-values less than 0.01}
\label{tab:dem_corr}
\begin{tabular}{ll}
Domain               & Corr. \\ \hline
nature.com              & 0.71        \\
medrxiv.org             & 0.58        \\
webmd.com               & 0.54        \\
thelancet.com           & 0.53        \\
bmj.com                 & 0.50        \\
thelancet.com           & 0.48        \\
wbur.org                & 0.45        \\
bostonglobe.com         & 0.41        \\
bbc.com                 & 0.53        \\
nejm.org                & 0.52        \\
amazon.com              & 0.50        \\
theatlantic.com         & 0.46        \\
instagram.com           & 0.45        \\
pubmed.ncbi.nlm.nih.gov & 0.45        \\
ncbi.nlm.nih.gov        & 0.40       
\end{tabular}
\end{wraptable}
Related to the politicization of URLs is who is posting them. For domains with at least thirty instances and ten authors, there are 10 domains where less than half of the users who posted them are currently active - the other half being either suspended or banned (Table \ref{tab:url_users}). Six of these domains are news domains (several relatively popular). For many of these, the number of posts is far higher than the number of authors who posted them, suggesting they were posted multiple times by one author, which may have contributed to the suspensions or bans (although the posts and comments were not deleted or removed). Generally, of 248611 distinct author-URL pairs, in 7320 cases the author shared the URL with at least two separate subreddits - only about 3\% of the time, suggesting obvious spammers and bots are rare among active users.

\begin{table*}
\begin{tabular}{lrrrlrrl}
\toprule
{} &    \% active &  \% banned &  \% suspended &                    domain &  URL count &  authors count &      class \\
\midrule
0 &  14.0 &   6.9 &   79.2 &                   cutt.ly &    159 &       31 &      other \\
0 &  24.2 &   6.0 &   69.7 &                 linktr.ee &     99 &       35 &  computers \\
0 &  39.5 &   54.4 &  6.1 &              huffpost.com &    114 &       32 &       news \\
0 &  43.2 &   51.0 &   5.8 &            in.reuters.com &    206 &       34 &       news \\
0 &  45.4&   30.8 &   23.8 &  www.thegatewaypundit.com &    130 &       40 &       news \\
0 &  46.4 &   38.4 &   15.3 &         www.breitbart.com &    209 &       55 &       news \\
0 & 47.3 &   44.8 &   7.9 &              www.aier.org &    165 &       41 &      other \\
0 &  47.4 &   45.5 &   7.1 &               theweek.com &    310 &       81 &       news \\
0 &  48.1 &   47.4 &  4.5 &          sports.yahoo.com &    133 &       32 &       news \\
0 &  48.8 &   50.1 &  0.3 &              www.cebm.net &    340 &       78 &     science \\
\bottomrule
\end{tabular}
\caption{User statuses for domains with at least 30 uses and 10 authors, where fewer than half of authors are active. Percentages are percent of instances of URL where user has a given status.}
\label{tab:url_users}
\end{table*}

\subsection{Keywords}
Another method of analyzing content at scale is tracking keywords. We identify keywords by using the Fighting Words \cite{monroe2008fightin} method, which identifies words that differentiate two sets of strings. By comparing state subreddits, we can analyze major differences in regional conversations. Logically, most states have a lot of location-specific terms including cities or politicians. Additional terms such as 'vaccinated', and 'cvs' may suggest some topics differences, but overall most states seem to have similar discussion topics.

\begin{figure}
    \centering
    \includegraphics[width=0.22\textwidth]{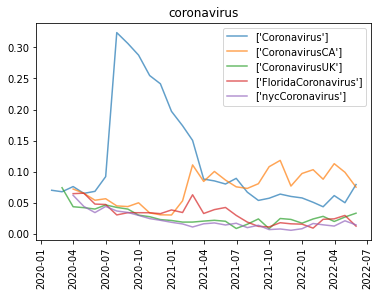}
    \includegraphics[width=0.22\textwidth]{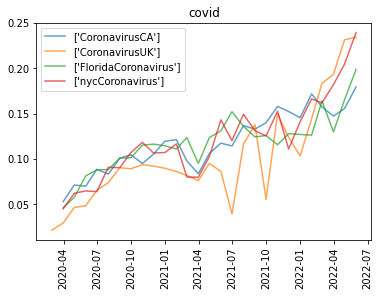}
    \includegraphics[width=0.22\textwidth]{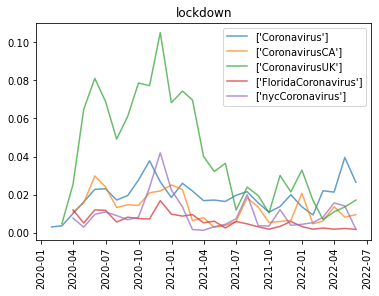}
    \includegraphics[width=0.22\textwidth]{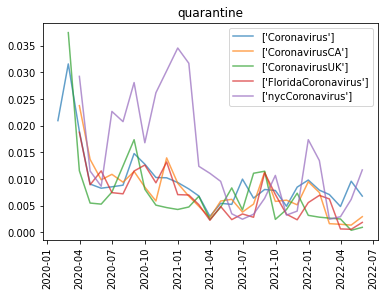}
    \caption{Percent of posts and comments that contain various terms across several subreddits over time. }
    \label{fig:keyterms}
\end{figure}

While we see that terms vary geographically, there is also a strong temporal aspect. We analyze the popularity of keyterms that are selected by doing a Fighting Words comparison between r\/nba and r\/Coronavirus to identify words specific to the COVID conversation (Fig.~\ref{fig:keyterms}). One interesting case the use of the terms 'coronavirus' and 'covid'. 
% The use of `coronavirus' spikes at the beginning. The term `covid', however, slowly increases in popularity throughout the pandemic, eventually overtaking the popularity of `coronavirus'. This may represent a change in terms as users shift how they represent the pandemic. 
The use of `coronavirus' spikes at the beginning, but `covid' slowly increases in popularity and eventually overtakes `coronavirus', reflecting a change in the dominant terminology.
`Lockdown' versus `quarantine' tells a different story. Both terms are popular during similar time periods, but `lockdown' shows up much more in non-US subreddits than `quarantine' does.

\section{Users Posting Content}

\begin{wraptable}{r}{3cm}
\begin{tabular}{ll}
Subreddits & Users \\ \hline
1 & 202266 \\
2 & 40911 \\
3 & 9669 \\
4 & 2737 \\
5 & 951 \\
6 & 352 \\
7 & 90 \\
8 & 49 \\
9 & 19 \\
10 & 7 \\
11+ & 19
\end{tabular}
\caption{Number of users who have participated in $n$ subreddits.}
\label{tab:num_subs}
\end{wraptable}

Our sample of COVID communities on Reddit included 260,000 active users, 22,000 banned users, and 16,000 suspended users. We focus our analysis on active users. For the purposes of this work, we only are able to see user interactions with a subreddit - either through posting or commenting - but not viewing or subscribing. Thus, we only consider which communities users interact with directly. We see that users that interact with more communities are much more active overall, posting and commenting far more than those with fewer communities. However, users also seem to have a `dominant' subreddit where they spend most of their time - and users have about 50\% of their interactions occur within this dominant subreddit regardless of how many subreddits they interact with. These additional communities also show a tendency to broaden their communities vertically before horizontally.

\begin{figure}
    \centering
    \includegraphics[width=0.4\textwidth]{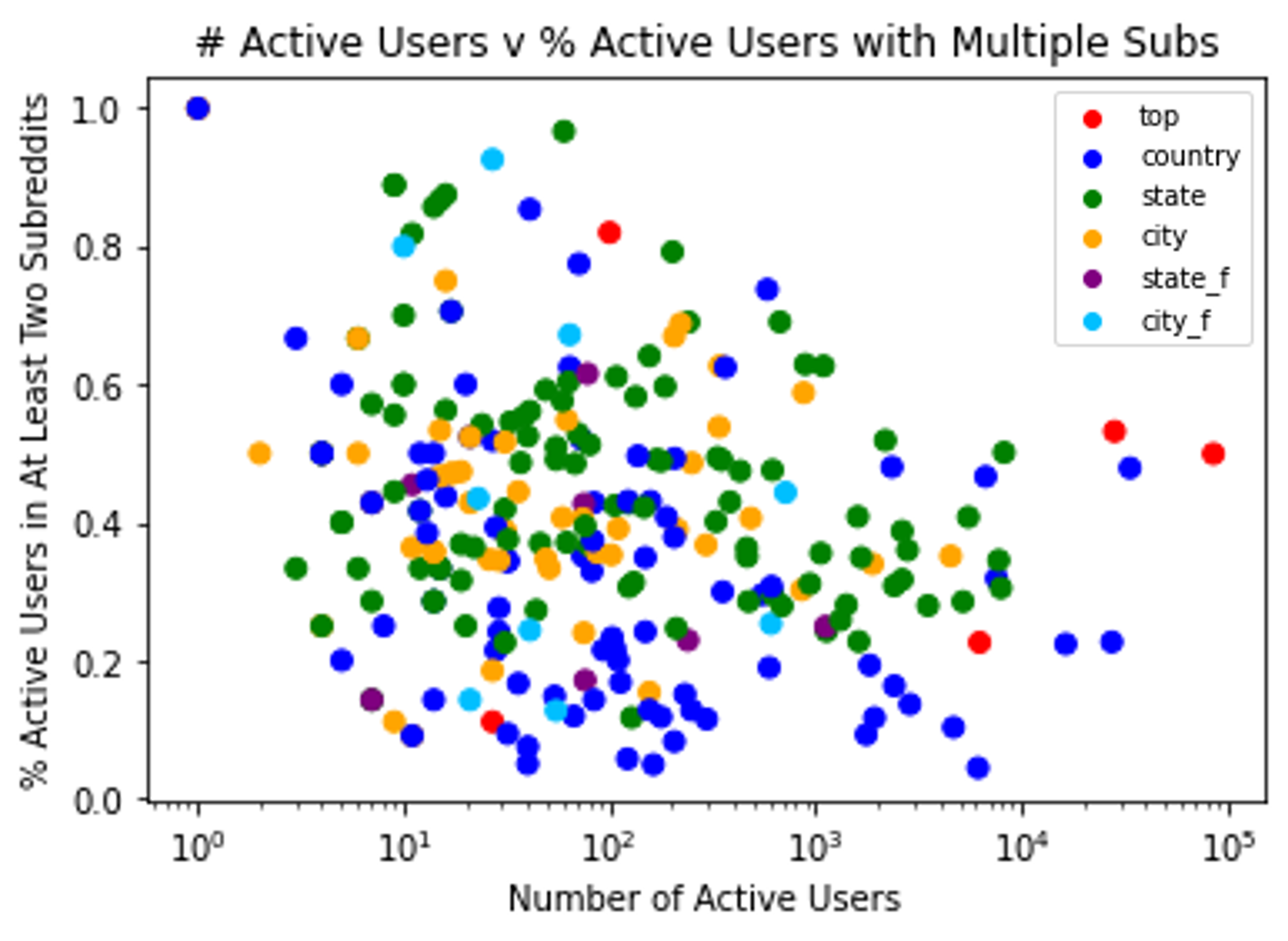}
    \caption{Total number of active users vs percentage of active users active in at least one other subreddit for each subreddit.}
    \label{fig:active_overlap}
\end{figure}

The vast majority users are part of only one subreddit, and only a handful of users are part of more than ten subreddits (Table \ref{tab:num_subs}). For those who are active in only one subreddit, 86224 are part of a country subreddit, 61013 part of a top subreddit, 45732 part of a state subreddit, 7127 part of a city subreddit and the rest part of subreddits for non-US states and cities (\~2000). Interestingly, if we look at how much of a subreddit's user-base also participates in other subreddits, we see a very large percentage across the board. The number of subreddits with very large overlap percentages decreases as we look at larger and larger subreddits, but not nearly as much as we would expect if we scrambled user membership via a null model with bipartite randomization (Fig.~\ref{fig:active_overlap}). This both shows that users with just one subreddits are more likely to participate in large subreddits (possibly because they are easier to find, or because smalls subreddits to not satisfy the need of a user alone), and that there is significantly more user overlap than would be predicted at random in general.

\begin{figure}
    \centering
    \includegraphics[width=0.4\textwidth]{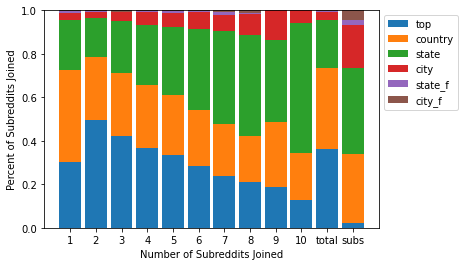}
    \caption{The percentage of subreddits participated in at each level, across all users that have participated in $x$ subreddits. The total bar is an overall percentage for all users, and the sub bar is the percentage of subreddits at each level.}
    \label{fig:sub_levels_stacked}
\end{figure}

If we compare this to users that participate in multiple subreddits, we can see that this is a higher percentage of country subreddits and lower percentage of top subreddits. It is notable that there are more country subreddits than top subreddits (Fig.~\ref{fig:sub_levels_stacked}). In fact, it seems that, apart from those with only one subreddit, the overall percentage of country subreddits is fairly stable. This is not true for top subreddits and state subreddits. Since there are only three top subreddits, we must expect that percentage to naturally decline, as we in fact do, but we also see a large sudden jump in the percentage of top subreddits when go from one subreddit to two subreddits. This means a top level subreddit is very often a second choice of subreddit for a user. The large increase in the percentage of state subreddits suggests that most users who participate in multiple subreddits are looking for additional communities at the state level - we now must ask why.

\begin{figure}
    \centering
    \includegraphics[width=0.4\textwidth]{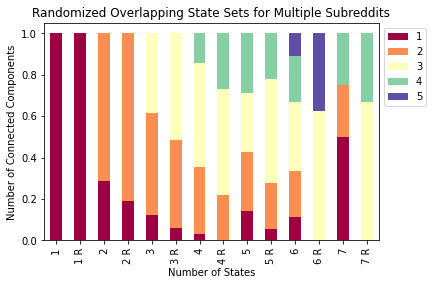}
    \caption{For users with $x$ non-isomorphic state subreddits, how many distinct groups of geographically adjacent states these states form. The bars with "R" have had the state-user pairs randomized as a null model.}
    \label{fig:overlapping_states}
\end{figure}

If we look at where the states are geographically located (Fig.~\ref{fig:overlapping_states}), we can see that in most cases users are very likely to choose states that are geographically close, more so than if we randomized user-subreddit pairs. This could express a desire for regional subreddits that are not constrained by specific state boundaries.

Moving beyond just the horizontal specialization of states, this level notation allows us to ask questions about the implicit vertical hierarchy. If a user is part of a city subreddit, are they also part of the relevant state and country subreddit? A user who is in a city subreddit is in the corresponding state or country subreddit about 25\% of the time. The number is similar for state subreddits. However, far fewer of those with city and state subreddits are part of a top subreddit (Table \ref{tab:hierarchy}).

\begin{wraptable}{l}{5.7cm}
\begin{tabular}{lrrrr}
\toprule
{} &       top &   country &     state &      city \\
\midrule
top     &       - &  0.0644 &  0.0459 &  0.0630 \\
country &  0.9356 &       - &  0.2200 &  0.2582 \\
state   &  0.9541 &  0.7800 &       - &  0.2406 \\
city    &  0.9370 &  0.7418 &  0.7595 &       - \\
\bottomrule
\end{tabular}
\caption{Fraction of instances of being in a \{column\} subreddit where one is also in the corresponding \{row\} subreddit for the upper triangle, and the complement for the lower triangle.}
\label{tab:hierarchy}
\end{wraptable}

\begin{figure}
    \centering
    \includegraphics[width=0.4\textwidth]{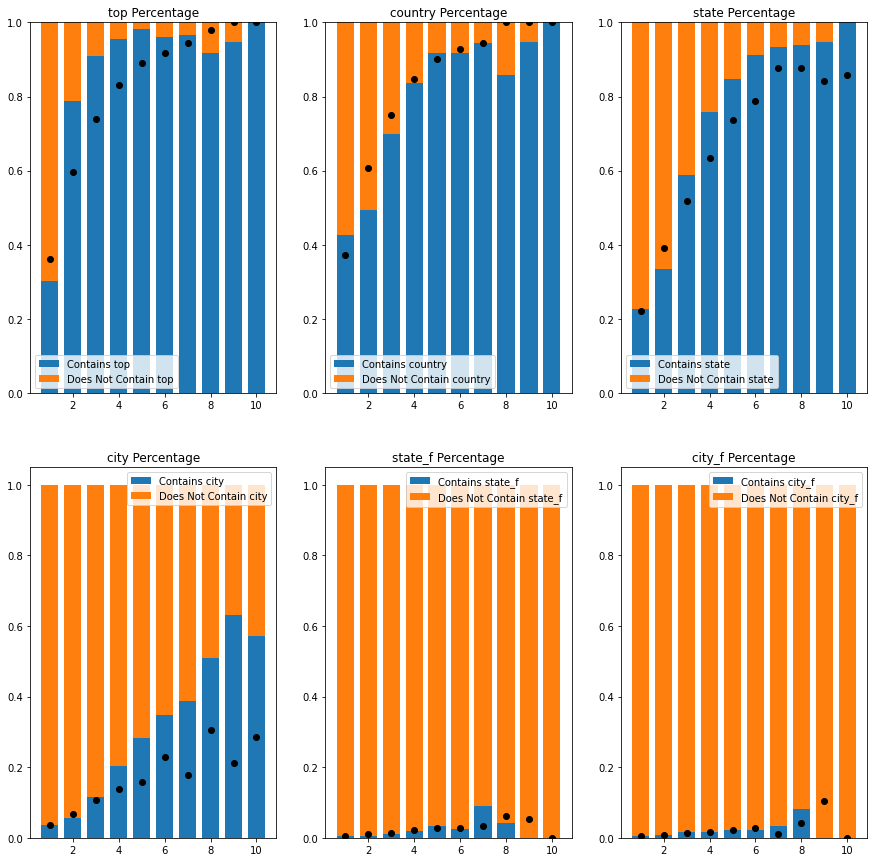}
    \caption{Number of users with $x$ subreddits where their subreddit list contains or does not contain at least one subreddit of a given level. The black dots represent a null model of user-subreddit pair randomization.}
    \label{fig:size_v_perc}
\end{figure}

We then again turn to look at the differences between users that are part of few or many subreddits (Fig.~\ref{fig:size_v_perc}). In this case, we can see that the percentage of users who have a subreddit of a particular level is somewhat dependent on the number of subreddits they are part of total. The likelihood of being part of a top level subreddit jumps precipitously from one subreddit to two subreddits, reflecting this tendency to vertical diversity. Similarly, both country and state percentages seem to increase in a roughly exponential way, but city percentage appears more linear. Also, while country and state percentages are somewhat close to our random baseline, both top and city subreddits deviate from those baselines, suggesting that our top and bottom levels are fundamentally different somehow from middle levels.

If we take subreddits to be nodes and connect those nodes if they share some minimum percentage of users, then we have a small simple graph of subreddit overlaps. This graph has one giant component, with a highly connected core and a few periphery nodes. However, if we increase the minimum percentage, we get a more focused view of the core of this graph. Another concrete way to testing the geographic influence on community membership, is to ask if the number of geographically significant connections remains the same for randomized graphs. We use both the configuration model and the k-core model \cite{van2021random} as null models and check how many connections there are between adjacent and non-adjacent states. 

For adjacent states, the true graph has about 1.9\% of edges as edges between adjacent states. For the configuration model, the average drops to 1.0\%, but for the k-core model we average 1.6\% - close to the original. This suggests that much of the adjacent state connections are reflected in the core structure of the graph - small clusters of interconnected states. However, for non-adjacent states the average percentage is much higher for both null models (13 \% and 15\%, respectively) than the original (3\%), which reflects a huge geographic influence in state connections - far fewer connections between non-adjacent states than we would expect at random. Rather, these edges may instead be part of more vertical connections than expected at random. Finally, for isomorphic states, fewer than 1\% of edges are edges between isomorphic subreddits for both random models, whereas nearly 8\% of edges are edges between isomorphic subreddits in the real graph. These measurements reinforce the fact that geography is highly influential in the communities users are a part of.

\begin{figure}
    \centering
    \includegraphics[width=0.4\textwidth]{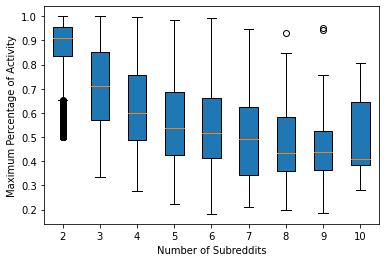}
    \caption{Maximum percentage of activity in one subreddit for those in $x$ subreddits, this shows percent activity in a user's dominant subreddit.}
    \label{fig:activity_box}
\end{figure}

Membership merely requires one action by a user in a subreddit. We can also analyze additional interactions, mainly posting and commenting. One natural question is whether or not users tend to have a 'dominant' subreddit - one subreddit with which they interact the most. By looking at the percentage of activity in the dominant subreddit for users in various numbers of subreddits (Fig.~\ref{fig:activity_box}), we can see the answer is generally yes. The median percentage is well above what we would expect if activity were evenly distributed no matter how many subreddits a user is part of. In fact, the percentage of activity in a user's dominant subreddit does not decrease much once a user has five subreddits, hovering at around nearly 50\%.

\begin{figure}
    \centering
    \includegraphics[width=0.4\textwidth]{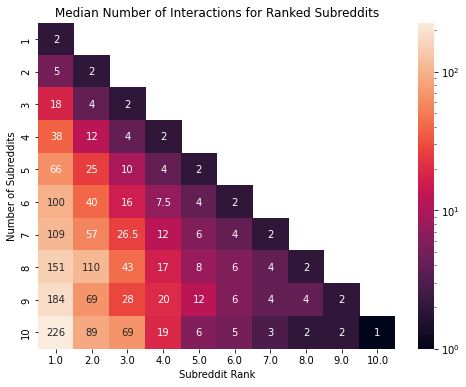}
    \includegraphics[width=0.4\textwidth]{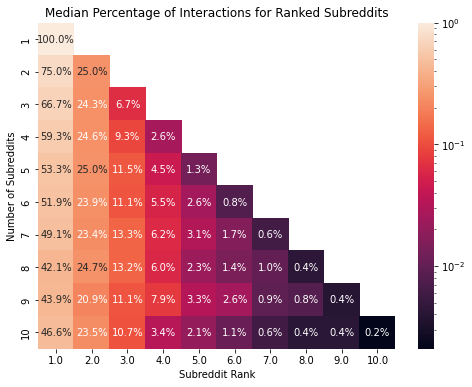}
    \caption{Median number of interactions (top) and percentage of interactions (bottom) with the $x^{\rm th}$ most interacted-with subreddit for users that interact with $y$ subreddits.}
    \label{fig:activity_heatmap}
\end{figure}

\begin{figure}
    \centering
    \includegraphics[width=0.4\textwidth]{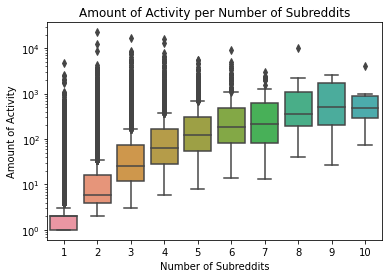}
    \caption{Distribution of number of total interactions for users in $x$ subreddits.}
    \label{fig:activity_per_sub}
\end{figure}

If we look at this distribution of activity another way (Fig.~\ref{fig:activity_heatmap}), we can see that the median percentage of activity for the top ranked subreddit is very high for all numbers of subreddits, but while the percentage decreases as the number of subreddit increases, the total number of interactions actually goes up. That is, users with many subreddits are generally more active in their top subreddits. This increase in counts would suggest that users with more subreddits are generally more active than users with fewer subreddits. This bears out in fact. Generally, the more subreddits one interacts with, the more activity one has overall (Fig.~\ref{fig:activity_per_sub}). In fact, this increase in activity outweighs the fact that the activity is spread across more subreddit, and generally more subreddits means more activity.

\begin{figure}
    \centering
    \includegraphics[width=0.22\textwidth]{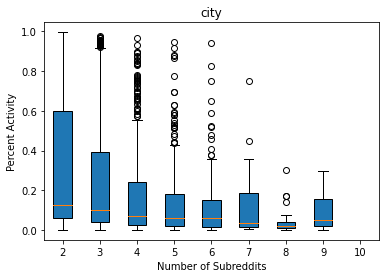}
    \includegraphics[width=0.22\textwidth]{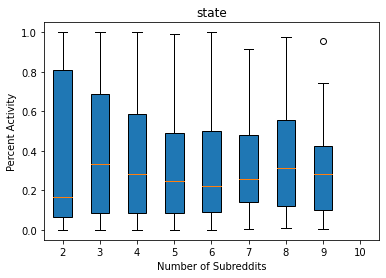}
    \includegraphics[width=0.22\textwidth]{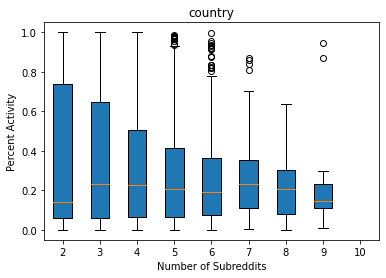}
    \includegraphics[width=0.22\textwidth]{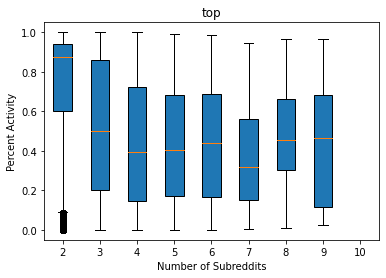}
    \includegraphics[width=0.22\textwidth]{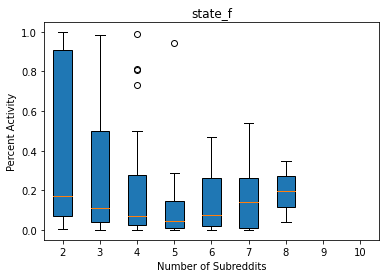}
    \includegraphics[width=0.22\textwidth]{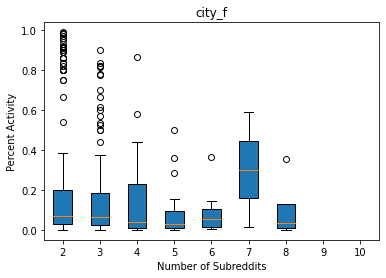}
    \caption{Percentage of activity in a given level for users with at least one subreddit in that level and $x$ subreddits in total.}
    \label{fig:boxplots}
\end{figure}

Looking at activity through the lens of vertical levels of geographic organization --- across different spatial scales --- one interesting point is that some levels have more activity than others (Fig.~\ref{fig:boxplots}); both top and country have generally high levels of activity, for instance. Also, while one might expect the percentage of activity to decrease as the number of subreddits increases (since more subreddits of other levels may be introduced), this is not strictly the case. The median percentage of activity remains consistent across the board. And while the 75th percentile does decrease somewhat, it does not decrease at all for top. In fact, for both state and country the median increases from 2 to 3 subreddits. This may suggest that users in multiple subreddits are generally more active in regional subreddits, but there is something mysterious about the fact that this fraction remains fixed regardless of the number of subreddits.

\section{Communities}
Reddit communities are user created, user moderated, and user organized. Thus, the creation and organization of these communities is a crowd-sourcing exercise, that describes how the larger Reddit community chose to organize the discussions about COVID. Nearly all communities were created within a short period, and these communities sometimes reference the same geographic areas - which we call \textbf{isomorphic} communities - but there is almost always one dominant community for a given location. Additionally, the creators of these subreddits tend to remain heavily involved in the community. While the conversation around COVID never died out during our time period, it did wax and wane. These fluctuations are often in line with three major waves of the epidemic: the initial wave, the Delta wave in mid 2021 and the Omicron wave in late 2021 and early 2022. 

\subsection{Subreddit Creation}
\begin{figure}
    \centering
    \includegraphics[width=0.4\textwidth]{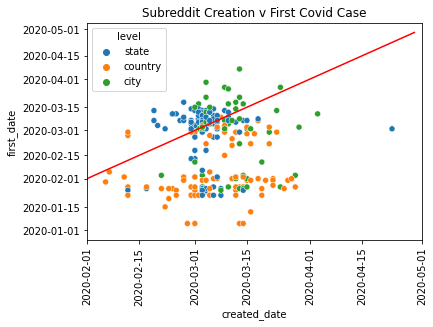}
    \caption{Date of creation of local subreddit vs date of first recorded COVID case. Subreddits above the line had a subreddit before the first case, those below had a COVID case before the subreddit was created.}
    \label{sub_creation}
\end{figure}

Overall, the vast majority of subreddits were created right around when COVID became a global concern - when it began to pop up in countries world wide. It was declared a pandemic on March 11, 2020. Most subreddits were created between March 1st and March 15th, slightly predating the pandemic designation. As can be seen in Figure \ref{sub_creation}, the first communities were related to countries, with states and cities closely behind. This mirrored the dates of first COVID cases, with most cities and states creating communities before the first local case, but after the first case in the country.

However, of the 5 top level subreddits, 3 were created in January, well before the first US COVID case and one in early February, predating almost all other subreddits. The final top level subreddits is r/Coronavirus, a co-opted subreddit first created in 2013 for general discussion on coronaviruses. This subreddit would later be designated the "official" COVID subreddit by Reddit and its initial purpose overturned. It is the most active COVID subreddit.

The next obvious question is - who created these communities? There are 249 distinct users who created the 281 identified subreddits. Most users create only one subreddit. 19 users created more than one, and only 6 created more than two. The most prolific creator, cryptodude1, created 8 subreddits for 8 different states - all on March 1st and 4th. Only one of those subreddits became popular. The next most prolific, askcoronavirus, created 5 variations of subreddits for Michigan, mostly on February 24th, with only one that had any real activity. This pattern continues, with most creators who create more than one subreddit creating state subreddits.

\begin{figure}
    \includegraphics[width=0.4\textwidth]{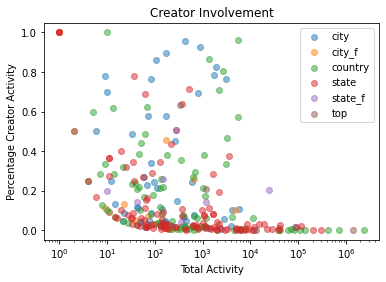}
    \caption{Percentage of activity in a subreddit by a creator versus the total amount of activity in that subreddit.}
    \label{fig:creator_involvement}
\end{figure}
Creator involvement post-creation varies wildly (Fig.~\ref{fig:creator_involvement}). While there is a natural decline in the amount of creator involvement as the size of the subreddit increases, among smaller subreddits (even into the thousands of posts/comments), some creators remain highly involved, accounting for almost all activity in the subreddit. For most locations, a subreddit is not considered "popular" until tens of thousands of interactions, however even among these popular subreddits a few have remarkably involved creators. A few subreddits (including r/Coronavirus) are excluded from this since we could not easily identify the creator. 

\omt{
\begin{figure}
    \centering
    \includegraphics[width=0.4\textwidth]{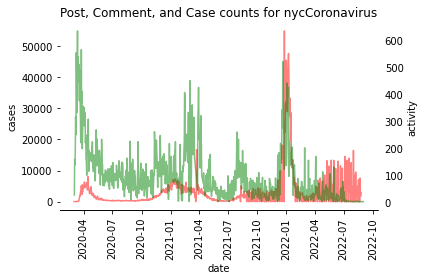}
    \includegraphics[width=0.4\textwidth]{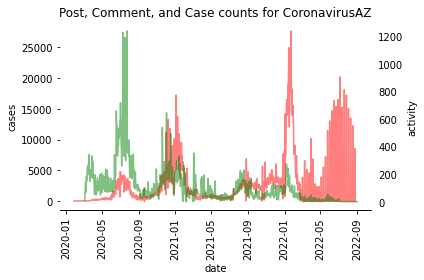}
    \caption{COVID cases versus amount of activity. Red is number of cases per day, green is number of posts and comments per day.}
    \label{fig:covid_activity}
\end{figure}
}

\subsection{Isomorphic Subreddits}
\begin{figure}
    \includegraphics[width=0.23\textwidth]{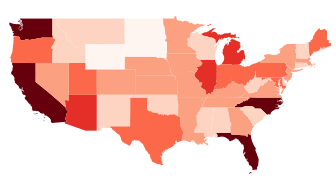}
    \includegraphics[width=0.22\textwidth]{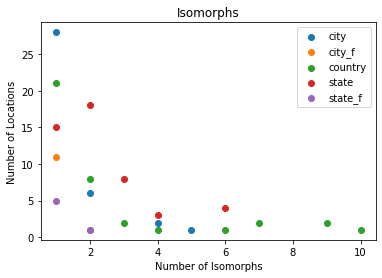}
    \caption{Most states in the US had at least two subreddits and up to six. Three states had none.}
    \label{fig:us_isomorphs}
\end{figure}

As mentioned previously, many locations had multiple subreddits. We call these "isomorphic" subreddits. These are most common for state subreddits, where only 15 states have one subreddit (and three have none). However, countries also frequently have isomorphs. The US has 10 isomorphic subreddits, and Canada and India each have nine, the UK and 'Europe' each have 7, and Australia has 6 (Fig.~\ref{fig:us_isomorphs}). These are the top subreddits with the most isomorphs. They also, notably, include a large portion of the English-speaking world.

Additionally, in most cases the first isomorphic subreddit is created within two weeks of the original subreddit, and often much sooner (Fig.~\ref{fig:creation_delay}). We did not include subreddits with explicit ideologies, only general COVID subreddits. Instead, most isomorphic subreddits are created by different users in close succession. If a subreddit has as least twice as much total activity as the next most popular subreddit for a location, we call it 'dominant'. Of 142 distinct locations, only 6 lack a dominant subreddit. For the 56 locations where there are at least two competing subreddits, in 35 instances the dominant subreddit is the first created, suggesting a bias towards initial creation.

\section{Conclusion}
Our analysis of the geographic specialization of online COVID communities allows for new insights into “horizontal” specialization across similarly-sized regions and “vertical” specialization into a hierarchy of communities. Against this backdrop, we analyze the content discussed in these communities in the form of URLs and keywords, the behavior of users in terms of which communities they participate in and how much they participate, and the self-organization of the online discussion into these communities.

While the COVID pandemic provided a unique opportunity to study these geographically specialized communities at a large scale, it is by no means the only place where this geographic specialization exists. Many of the themes we discuss in this work are applicable to other, non-COVID sets of communities and it would be interesting to understand the influence of the COVID topic. Similarly, this analysis is primarily US focused, and it would be interesting to produce analysis at similarly fine-grained resolution for the hierarchy of geographies in more non-US-based communities. Additionally, while this work focuses only on active participation of users, expanding the analysis to readership would provide additional context.

% \bibliographystyle{plain}
% \bibliography{works_cited}

\clearpage
\appendix
\section{Appendix}
\FloatBarrier

\begin{figure}
    \centering
    \includegraphics[width=0.5\textwidth]{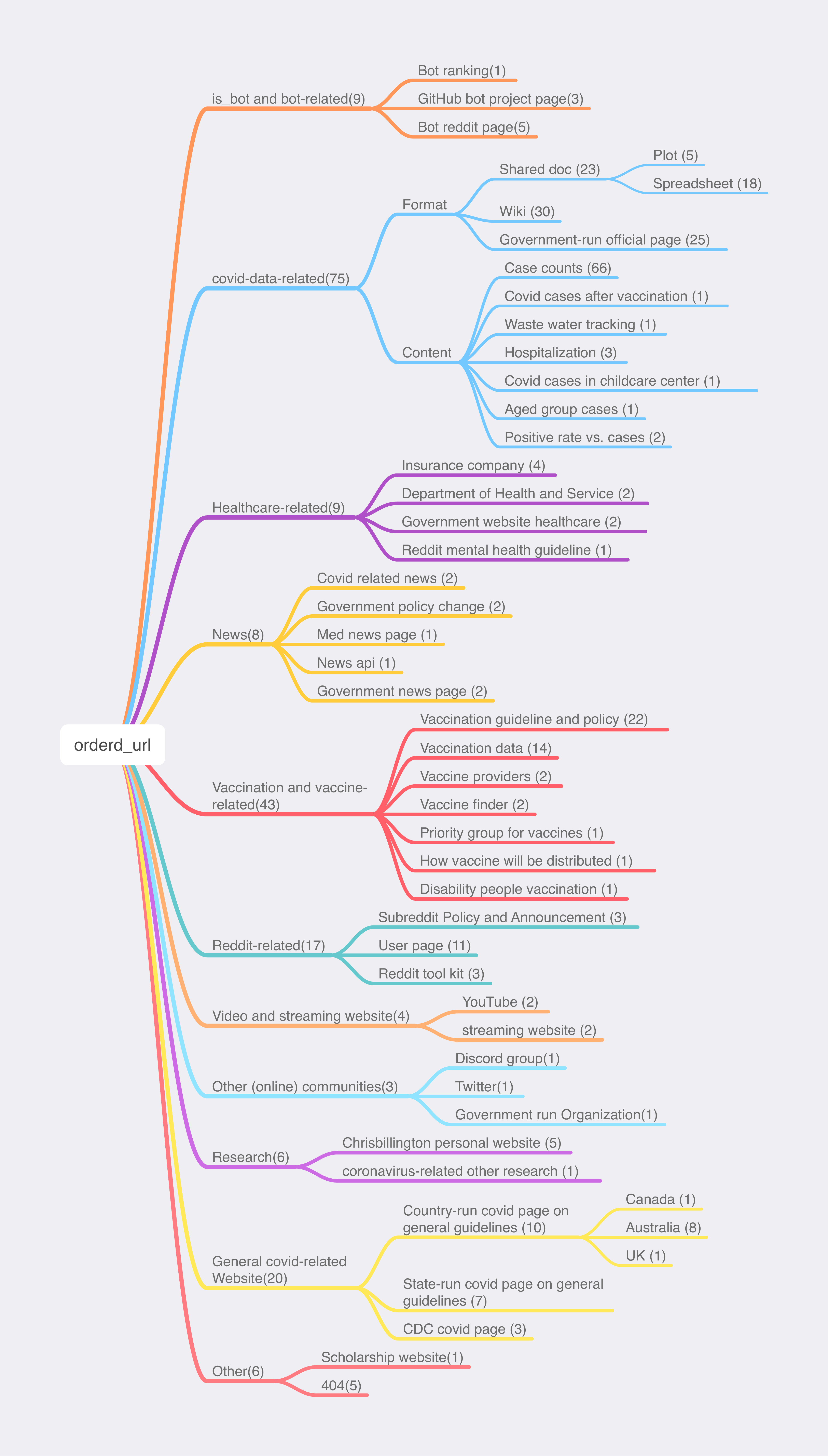}
    \caption{A classification tree for the top 200 most commonly shared URLs across all communities.}
    \label{fig:url_tree}
\end{figure}{}

\begin{figure}
    \includegraphics[width=0.5\textwidth]{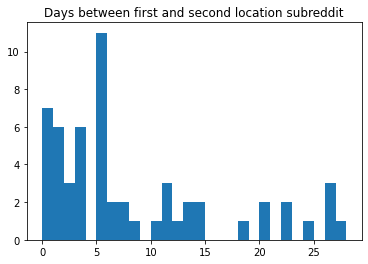}
    \caption{Histogram of the delay between creation of first and second subreddit for all locations with at least two subreddits.}
    \label{fig:creation_delay}
\end{figure}

\begin{figure*}
    \centering
    \includegraphics[width=0.45\textwidth]{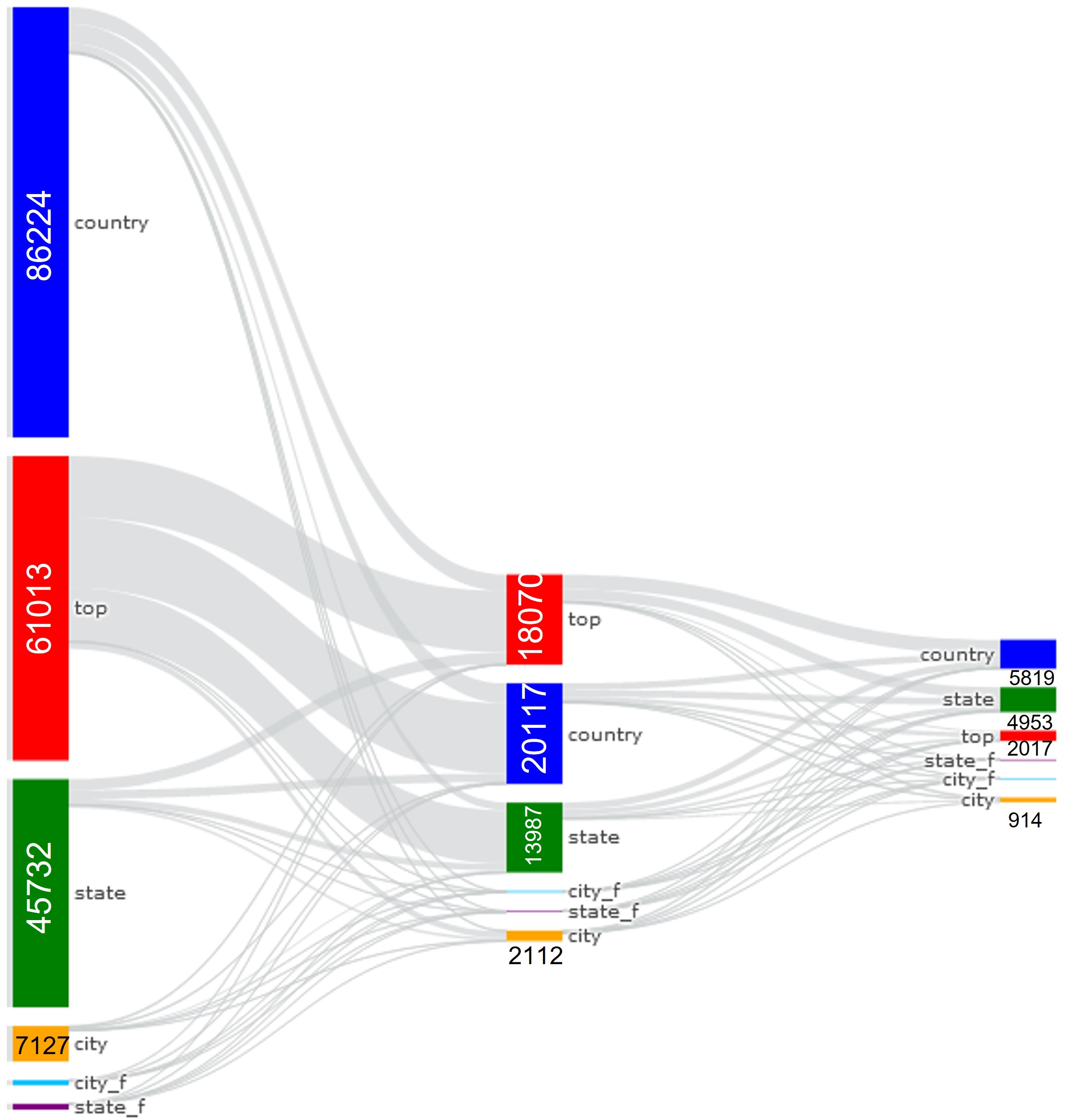}
    \caption{Sankey diagram of the levels of the first three subreddits participated in for all users. Note that most users begin with a country or state subreddit, but many of those never participate in a second subreddit. However, most users that start with a top subreddit, go on to participate in at least one additional subreddit, suggesting a strong inclination towards local community.}
    \label{fig:sankey}
\end{figure*}

\begin{figure*}
    \centering
    \includegraphics[width=0.55\textwidth]{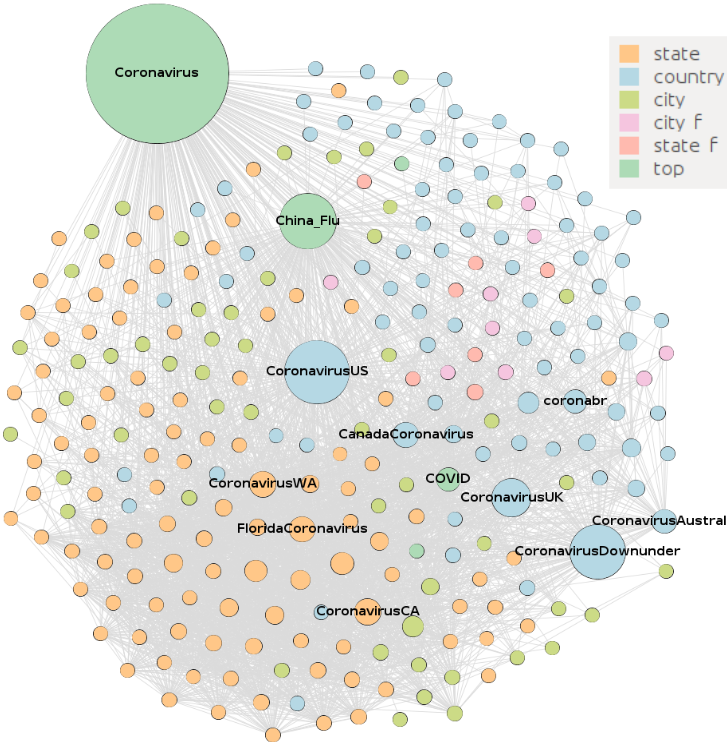}
    \caption{Subreddit network where each edge represents a user overlap of at least 10\% of the smaller subreddit. Node sizes are proportional to number of users in the community, and nodes are colored by level.}
    \label{fig:network}
\end{figure*}

\end{document}